\documentclass[preprint,12pt]{elsarticle}
\usepackage[version=4]{mhchem}
\usepackage{mathtools}
\usepackage{csquotes}
\usepackage{graphicx}
\usepackage{float}
\usepackage{dcolumn}
\usepackage{bm}
\usepackage{color}
\usepackage[table,xcdraw,dvipsnames]{xcolor}
\usepackage{multirow}
\usepackage{array}
\usepackage{booktabs}
\usepackage{siunitx}[=v2]
\usepackage[breaklinks]{hyperref}
\usepackage{amssymb}
\usepackage{amsmath}
\usepackage{fullpage}
\usepackage{diffcoeff}
\usepackage[frozencache, cachedir=.]{minted}
\usepackage[most, minted]{tcolorbox}
\usepackage{ulem}
\usepackage{xurl}

\DeclareMathOperator*{\argmin}{arg\,min}

\DeclarePairedDelimiter{\pqty}{(}{)}

\DeclarePairedDelimiter{\vqty}{\vert}{\vert}
\DeclarePairedDelimiter{\Bqty}{\lbrace}{\rbrace}
\DeclarePairedDelimiterX{\braket}[2]{\langle}{\rangle}{#1\,\delimsize\vert\,\mathopen{}#2}
\DeclarePairedDelimiterX{\mel}[3]{\langle}{\rangle}{#1\,\delimsize\vert\,\mathopen{}#2\,\delimsize\vert\,\mathopen{}#3}

\newtcblisting{pycode}{
  listing engine=minted,
  minted language=python,
  minted options={
      fontsize=\small,
      fontfamily=txtt,
      mathescape
  },
  colback=yellow!10,      
  sharp corners,              
  boxrule=0pt,                
  toprule=1pt,                
  bottomrule=1pt,             
  colframe=black,             
  enhanced,                   
  listing only,               
  left=10pt, right=10pt,      
  top=5pt, bottom=5pt
}

\newcounter{bla}

\journal{Computer Physics Communications}

\begin{document}
\begin{frontmatter}



\title{\textsc{Pulgon-tools}: A toolkit for analysing and harnessing symmetries in quasi-1D systems}


\author{
Yu-Jie Cen,$^1$ Sandro Wieser,$^1$ Georg K. H. Madsen,$^1$ and Jesús Carrete$^{2,*}$\\
{\normalsize{}$^1$ \textit{Institute of Materials Chemistry, TU Wien, A-1060 Vienna, Austria}}\\
{\normalsize{}$^2$ \textit{Instituto de Nanociencia y Materiales de Aragón, CSIC-Universidad de Zaragoza,}}\\
{\normalsize{}\textit{E-50009 Zaragoza, Spain}}\\
{\normalsize{}$^*$ \href{mailto:jcarrete@gmail.com}{jcarrete@gmail.com}}
}

\begin{abstract}
\textsc{Pulgon-tools} is an open-source software package providing building blocks for the analysis and modeling of quasi-one-dimensional (quasi-1D)  periodic systems based on line-group theory. While mature libraries exist for space-group detection in three-dimensional crystals, an automated and structure-based identification of line groups has so far been lacking. We present software that integrates four complementary components within a consistent line-group framework: (i) structure generation, (ii) symmetry detection, (iii) irreducible representations (irreps) and character tables and (iv) harmonic interatomic force constants (IFCs) correction.  This paper introduces the general code structure and several examples that illustrate some relevant applications of the program. 

\noindent \textbf{PROGRAM SUMMARY}

\begin{small}
\noindent
{\em Program Title: } \textsc{Pulgon-tools}                                          \\
{\em CPC Library link to program files:} (to be added by Technical Editor) \\
{\em Developer's repository link:} \href{https://github.com/pulgon-project/pulgon_tools/releases}{\url{https://github.com/pulgon-project/pulgon_tools/releases}} \\
{\em Licensing provisions:} Apache 2.0  \\
{\em Programming language:} Python                                   \\
{\em Nature of problem:} Quasi-1D periodic systems such as nanotubes and nanowires possess intrinsic symmetries described by line groups, which cannot be identified by existing three-dimensional space-group tools. No widely available software library currently enables automated line-group symmetry detection directly from atomistic structures. Furthermore, tools for generating quasi-1D structures consistent with line-group symmetry, computing irreps and character tables, and enforcing physical invariance conditions on harmonic interatomic IFCs in such systems are lacking.\\
{\em Solution method:} \textsc{Pulgon-tools} addresses these challenges through four complementary modules. Structure generation is supported via two approaches: a general symmetry-based construction that builds a periodic structure from a set of generators and an atomic motif, and a chiral roll-up method that generates \ce{MoS2}-type nanotubes directly from chiral indices. Symmetry detection proceeds in two stages: first, the generalized translational group is identified by testing the screw and glide symmetry, and the axial point group is determined by analyzing rotational and mirror symmetries. Irreps and character tables are constructed analytically following the literature, with each irrep labeled by physically meaningful quantum numbers. IFC correction enforces translational and rotational invariance conditions on second-order IFCs via constrained quadratic optimization. \\
   \\

\end{small}
   \end{abstract}
\end{frontmatter}

\section{Introduction}
Symmetry plays a central role in computational physics, providing essential information for understanding the structural, vibrational, electronic, and topological properties of crystalline systems, ensuring the physical correctness of results, and accelerating calculations \cite{dresselhaus2007group,el2008symmetry}. In atomistic simulation and data analysis, specifically,  the identification of crystal symmetry is a fundamental step, its usefulness ranging from reducing the computational cost of Brillouin-zone sampling \cite{PhysRevB.13.5188} to enabling symmetry-based classification of physical states \cite{PhysRevB.90.115438, Rousseau1981, PhysRevB.65.045418}. For three-dimensional periodic systems, this task is now well supported by mature and widely adopted software like \textsc{sgroup}\cite{Yanchitsky_CPC01} and \textsc{spglib}\cite{Togo31122024}
, which provides robust and automated space-group symmetry detection for atomistic structures. Furthermore, the \textsc{Bilbao Crystallographic Server} \cite{Bilbao} is a feature-rich interactive online platform for crystallographic symmetry analysis that includes irreps, group–subgroup relations for the space groups, plane groups, layer groups and rod groups, and many other capabilities, but does not cover line-group symmetries.  \textsc{FINDSYM} \cite{Stokes:zm5027} and \textsc{AFLOW Online} \cite{curtarolo:art104} are symmetry-detection websites that can also automatically determine the three-dimensional space group of given atomistic structures. \textsc{PyXtal} \cite{pyxtal} focuses on the generation and enumeration of crystal structures based on space-group symmetry, enabling symmetry-constrained structure construction. \textsc{ISOTROPY} \cite{StokesISOTROPY} is a group-theory-based software suite for crystallographic symmetry analysis, incorporating tools for isotropy subgroups and symmetry-adapted distortions, space groups and irreducible representations, and the study of structural phase transitions. 

In contrast, quasi-1D periodic systems, such as nanotubes and nanowires, have received comparatively less attention from the perspective of symmetry-detection tools. Although these systems are often embedded in three-dimensional simulation cells for computational convenience, their intrinsic symmetry is not governed by three-dimensional space groups. It is instead described by line groups \cite{damnjanovic2010line}, which consist of a generalized translational group combined with an axial point group. Line groups form a complete and well-established classification scheme for quasi-1D crystallographic symmetries and cannot be regarded as a simple subset or dimensional reduction of three-dimensional space groups. 

Unlike translational groups in space groups, the generalized translational group can include screw rotations and glide reflections along the periodic axis. Furthermore, unlike point groups in space groups, axial point groups can contain rotational operations with arbitrarily subdivided angles around the periodic direction, reflecting the reduced translational constraints in one-dimensional systems \cite{damnjanovic2010line,reich2002carbon}. These features lead to differences in both the mathematical structure of the symmetry groups and the algorithms required for their identification. Applying existing space-group software, such as \textsc{spglib} \cite{Togo31122024}, to these systems generally yields only low-symmetry space groups, because these unique symmetry operations can not be identified.

Software that deals with line groups is far less abundant. The Wolfram-Mathematica-based \textsc{POLSym} \cite{Milosevic1996_POLSym, Milosevic1994_thesis} is a tool implementing several features to deal specifically with quasi-1D systems from the point of view of line groups. In particular, it implements the modified-group-projector technique and can generate line group representations, invariant polynomials, Clebsch-Gordan coefficients and the quantities required for several lattice-dynamical operations. More recent updates are reported to include code to deal with optical spectra and magnons, among other improvements \cite{PolSYM2}. There is therefore a partial overlap with the feature set of the software described here. Unfortunately, the \textsc{POLSym} download links at \url{https://www.nanolab.rs} seem to be broken as of this writing, and despite extensive searches we have not been able to find any viable alternative download route, leading us to conclude that \textsc{POLSym} is no longer publicly available. Nevertheless, and based on the remaining documentation, \textsc{POLSym} is targeted at interactive use and does not provide an automated pipeline for symmetry detection from arbitrary atomistic structures.

Among first-principles packages, the \textsc{CRYSTAL} program \cite{Erba2023} natively supports some quasi-1D systems and exploits their helical symmetry to reduce the cost of self-consistent calculations. For structure construction, it can generate helically periodic systems from a motif and a helical operation specification, and can roll-up an arbitrary two-dimensional parent layer into a nanotube over a range of chiralities \cite{https://doi.org/10.1002/jcc.21370,Marana2021}. It is effectively restricted, however, to helical geometries belonging to the first family of line groups, and does not supply the additional mirror, glide or $U$-axis operations that characterize more general structures. \textsc{CRYSTAL} also supplies a strategy to obtain phonon spectra that respect helical symmetry, but its significant computational cost, restricting it to smaller monomers \cite{nano15070505}.

Related approaches that exploit the irreps of helical line groups to reduce quasi-1D problems to single helical cell have also been developed, including the revised periodic boundary conditions within a tight-binding framework \cite{PhysRevB.84.155431} and a real-space helical DFT formulation that establishes the existence and completeness of helical Bloch waves \cite{BANERJEE2021104515}.

Thus, despite the growing importance of quasi-1D materials in both fundamental research and technological applications \cite{BALANDIN202274,saito1997nanotube,Carrete_PCCP19,shen2010one}, to the best of our knowledge there is currently no widely available software library that enables the detection of line-group symmetries directly from atomistic structures.
To address this gap, we are introducing \textsc{Pulgon-tools}, an open-source software package specifically designed for symmetry analysis of quasi-1D periodic systems. The primary goals of \textsc{Pulgon-tools} are to determine the complete set of symmetry operations of a given structure, including its generalized translational group $Z$ and axial point group $P$, to identify the corresponding line group type, and to provide access to its irreps and their characters. In addition to symmetry detection, the software also provides multiple symmetry-resolved functionalities, including two complementary approaches for quasi-1D structure construction and capabilities for second-order-IFC correction for vibrational calculations. We have previously used \textsc{Pulgon-tools} to study thermal transport across defect-laden segments in double-layer \ce{WS2-MoS2} nanotubes employing a symmetry-adapted Green's-function technique achieving an irrep-resolved description \cite{Cen2026}. We note that \textsc{Pulgon-tools} targets commensurate quasi-1D systems, whose symmetry is compatible with a finite primitive translational period along the axis. Such systems can still be treated as three-dimensional periodic structures by most programs and libraries in the large ecosystem built around ab-initio calculations. Such systems admit a description in terms of the linear axial wave number $k$. Incommensurate quasi-1D systems, which require bespoke calculators and helical wave numbers, fall outside of its present scope.

This paper presents the theoretical background, algorithmic design, and implementation details of \textsc{Pulgon-tools}, along with application examples. Section~\ref{sec:overview} provides an overview of the software architecture, describing its modular design and data flow. Section~\ref{sec:algorithm} details the four core functionalities of the package and their underlying algorithms. Section~\ref{sec:examp} illustrates the application of these modules to representative quasi-1D systems. Finally, section~\ref{sec:sum} concludes with a discussion of the capabilities, limitations, and possible future developments of the software.

\section{General structure of \textsc{Pulgon-tools}}
\label{sec:overview}
\textsc{Pulgon-tools} is designed as a modular software package for the analysis and exploitation of line-group symmetries in quasi-1D periodic systems. 
From a software-engineering perspective, it follows a pipeline-oriented design, where each module addresses a distinct stage of the symmetry workflow. They are logically independent but connected through well-defined data interfaces, allowing them to be used either independently or as part of a complete analysis sequence.

\begin{figure}[H]
    \centering
    \includegraphics[width=.9\linewidth]{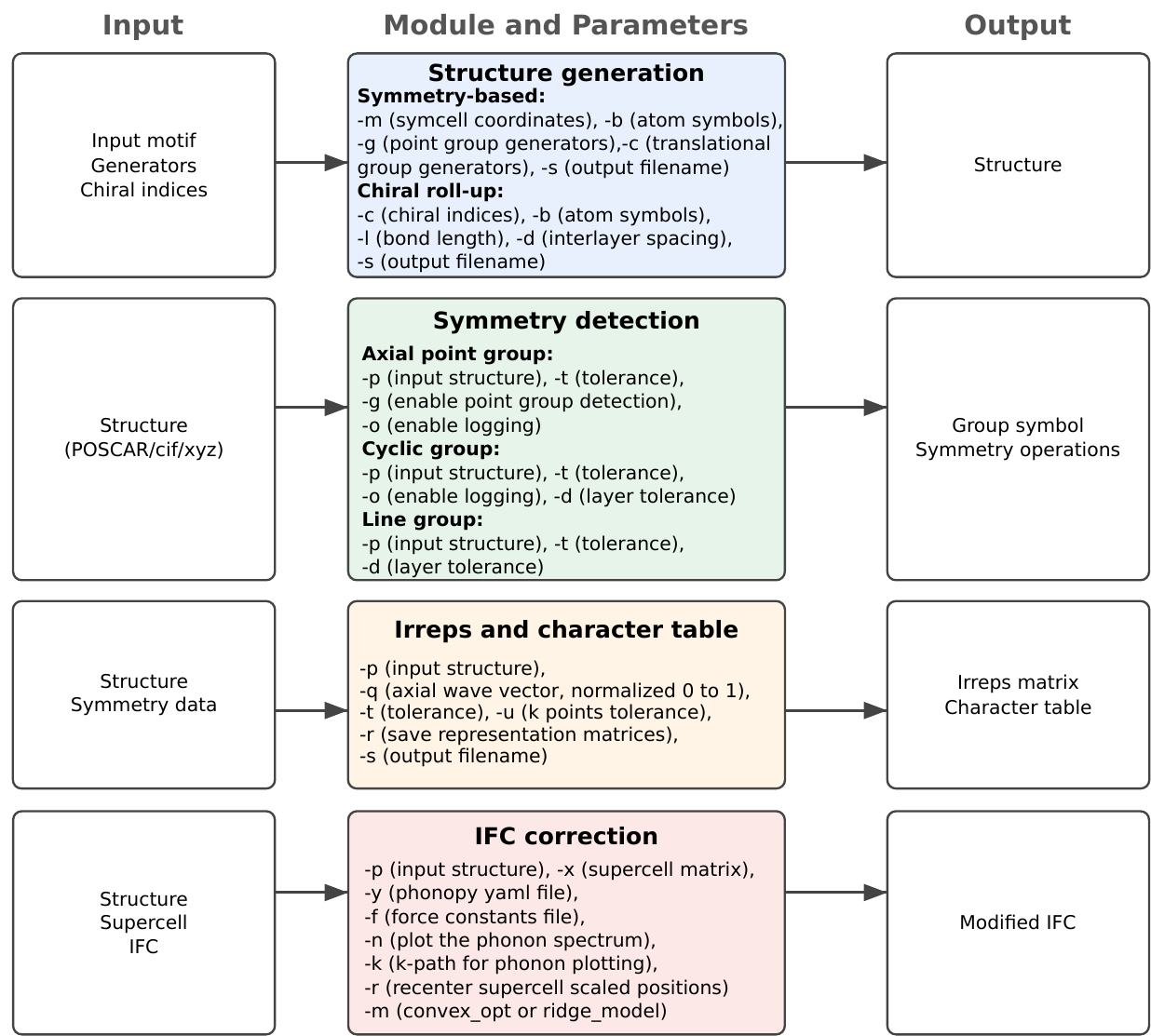}
    \caption{Data flow within \textsc{Pulgon-tools}.}
    \label{fig:in-output}
\end{figure}

Fig.~\ref{fig:in-output} summarizes the modular organization and data flow of \textsc{Pulgon-tools}. The architecture can be divided into four main components: structure generation, symmetry detection, irrep-and-character-table query and IFC correction.

\subsection{Structure-generation module}

The structure-generation module provides two complementary approaches for constructing quasi-1D systems.  
The first is a general symmetry-based construction, in which a quasi-1D structure is generated directly from the specified line-group generators and a minimal atomic motif. The monomer is then constructed by applying axial point group operations to the initial motif, and generalized translational operations are subsequently applied to generate the full periodic structure. The second approach is a chiral roll-up construction for \ce{MX2}-type nanotubes (one transition-metal atom \ce{M} and two chalcogenide atoms \ce{X} per formula unit, with \ce{MoS2} being the most common example), in which the structure is defined by chiral indices $(n, m)$ of a three-layer two-dimensional hexagonal parent lattice. The algorithm determines the corresponding helical line-group parameters and maps the planar structure to cylindrical coordinates. The final structure is constructed in a manner consistent with the underlying line-group symmetry.
Both approaches return standard atomistic structures that can be verified in the symmetry-detection module. 

\subsection{Symmetry-detection module}

The symmetry-detection module is used to determine the complete line-group symmetry of a given quasi-1D structure. The detection procedure is organized into two stages: generalized-translational-group identification, which determines the primitive translation along the periodic direction and detects possible screw or glide components; and axial-point-group identification, which classifies rotational, improper rotational, mirror, and dihedral symmetries around the periodic axis. These two stages are implemented as separate operations on the same structural input and their respective outputs are combined to determine the final line-group classification and the complete set of symmetry operations. 

In total, there are 13 line-group families, classified according to the combinations of generalized translational group $Z$ and axial point group $P$, as summarized in Table~\ref{tab:line_group} following the classification of Damnjanovi\'c and Milo\v{s}evi\'c \cite{damnjanovic2010line}. The resulting symmetry object serves as the input for irreps-and-character-table module.

\begin{table}[H]

\centering
\caption{The 13 line-group families and their defining symmetry components. $Z$ denotes the type of generalized translational group: pure translation $T$, screw $T_Q$, or glide $T_C$. $P$ denotes the axial point group. Notation adapted from table~2.2 in Ref.~\citenum{damnjanovic2010line}, after fixing a misprint for family $5$.}

\label{tab:line_group}

\begin{tabular}{c c c cc}

\toprule
Family & $Z$ & $P$ 
& \multicolumn{2}{c}{International symbol} \\\midrule
1  & $T_Q$ & $C_n$     
& \multicolumn{2}{c}{$\mathbf{L}q_p$} \\

& & & \cellcolor{black}{\small \textcolor{white}{$n$ even}} & \cellcolor{black}{\small \textcolor{white}{$n$ odd}} \\
2  & $T$   & $S_{2n}$  
& \cellcolor{lightgray}{$\mathbf{L}(\overline{2n})$} 
& \cellcolor{lightgray}{$\mathbf{L}\overline{n}$} \\

3  & $T$   & $C_{nh}$  
&  \cellcolor{lightgray}{$\mathbf{L}(2n)$} 
&  \cellcolor{lightgray}{$\mathbf{L}n/m$} \\

4  & $T_Q$ & $C_{2nh}$ 
& \multicolumn{2}{c}{$\mathbf{L}(2n)_n/m$} \\
& & & \cellcolor{black}{\small \textcolor{white}{$q$ even}} & \cellcolor{black}{\small \textcolor{white}{$q$ odd}} \\
5  & $T_Q$ & $D_n$     
& \cellcolor{lightgray}{$\mathbf{L}q_p22$} 
& \cellcolor{lightgray}{$\mathbf{L}q_p2$} \\

& & & \cellcolor{black}{\small \textcolor{white}{$n$ even}} & \cellcolor{black}{\small \textcolor{white}{$n$ odd}} \\

6  & $T$   & $C_{nv}$  
& \cellcolor{lightgray}{$\mathbf{L}nmm$}
& \cellcolor{lightgray}{$\mathbf{L}nm$} \\

7  & $T_C$ & $C_{nv}$  
&  \cellcolor{lightgray}{$\mathbf{L}ncc$} 
&  \cellcolor{lightgray}{$\mathbf{L}nc$} \\

8  & $T_Q$ & $C_{2nv}$ 
& \multicolumn{2}{c}{$\mathbf{L}(2n)_n mc$} \\

& & & \cellcolor{black}{\small \textcolor{white}{$n$ even}} & \cellcolor{black}{\small \textcolor{white}{$n$ odd}} \\
9  & $T$   & $D_{nd}$  
& \cellcolor{lightgray}{$\mathbf{L}(\overline{2n})2m$}
& \cellcolor{lightgray}{$\mathbf{L}\overline{n}m$} \\

10 & $T_C$ & $D_{nd}$  
& \cellcolor{lightgray}{$\mathbf{L}(\overline{2n})2c$} 
& \cellcolor{lightgray}{$\mathbf{L}\overline{n}c$} \\

11 & $T$   & $D_{nh}$  
& \cellcolor{lightgray}{$\mathbf{L}n/mmm$} 
& \cellcolor{lightgray}{$\mathbf{L}(2n)2m$} \\

12 & $T_C$ & $D_{nh}$  
& \cellcolor{lightgray}{$\mathbf{L}n/mcc$} 
& \cellcolor{lightgray}{$\mathbf{L}(2n)2c$} \\

13 & $T_Q$ & $D_{2nh}$ 
& \multicolumn{2}{c}{$\mathbf{L}(2n)_n/mcm$} \\

\bottomrule
\end{tabular}
\end{table}

\subsection{Irreps and character table module}
\label{subsec:irreps}
Based on the identified line group, this module provides direct access to irreps and character table of all $13$ line-group families. Each irrep is uniquely labeled by a set of physically meaningful quantum numbers: the axial wave number $k\in(-\pi/L,\pi/L]$, the angular momentum index $m$ and parity labels $\Pi$. The allowed values and physical interpretation of these quantum numbers depend on the specific line-group family. A detailed account is given in Sec.~\ref{subsec:irrep}. 

If the input structure is a supercell, the module will transform it to primitive and detect the line group family. Whenever a user-specified wave vector $k$ is needed, it always corresponds to the first Brillouin zone of the primitive cell.

Note that the current implementation always works with an axial wave number $k$, i.e., in the $(k, m)$ convention for quantum numbers. This is suitable for commensurate systems like those treated by \textsc{Pulgon-tools}. An alternative convention $(\tilde{k}, \tilde{m})$ exists, using a helical wave number, valid for incommensurate systems as well \cite{damnjanovic2010line}.

\subsection{IFCs correction module}
The IFCs correction module enforces physical invariance constraints on second-order IFCs, which could have been generated by \textsc{Phonopy} \cite{phonopy-phono3py-JPCM, phonopy-phono3py-JPSJ}, hiPhive \cite{hiphive} or other software \cite{CASTEP} in advance. Our IFCs correction module operates as a post-processing stage to fix the error and maintain the conservation of linear and angular momentum.

The implementation is designed to handle large sparse systems efficiently and can be used independently of the symmetry-detection module. However, when symmetry information is available, it can be used to reduce redundancy and improve numerical stability.

\section{Functional implementation details}
\label{sec:algorithm}

\subsection{Quasi-1D-structure generation}

\subsubsection{General symmetry-based approach}  
In the general symmetry-based construction approach, quasi-1D structures are generated directly from the intended line-group symmetry. This method is fully general and applicable to systems with arbitrary atomic motifs. The procedure thus requires two inputs: a target line group, specified by its generators, and the atomic motif, defined in a cylindrical coordinate system $\pqty*{r,\phi,z}$  with the periodic axis aligned along $z$.

The construction proceeds as follows:

\paragraph{(a) Group expansion}
Starting from a set of generators of the axial point group (e.g., rotations, mirror planes, or improper rotations), the complete point-group symmetry operations are generated by iterative group closure using a standard group expansion algorithm \cite{holt2005handbook}. Starting from the generators $\Bqty*{g_{1},g_{2},...}$, all pairwise products are computed and any new elements are added to the group until no further elements are generated.

\paragraph{(b) Monomer construction}
The initial atomic motif is acted upon by all axial-point-group operations to create the monomer, which represents the smallest non-redundant unit that can be combined with the generalized translational group to create the structure. Duplicate atomic positions generated by symmetry operations are removed within a numerical tolerance.

\paragraph{(c) Construction of the full structure}
The one-dimensional periodicity is introduced through generalized translational operator $Z$. There are two types of $Z$: a screw operation combining an $Q$-fold rotation with a fractional translation $f$ along $z$
\begin{equation}
    Z_{screw}=\pqty*{C_{Q} \mid f}: \qquad \pqty*{r,\phi,z}\mapsto  \pqty*{r,\phi + \frac{2\pi}{Q},z+f} ,
    \label{eqn:screw}
\end{equation}

\noindent and a glide operation consisting of a vertical mirror reflection plus a half-period translation:
\begin{equation}
    Z_{glide}=\pqty*{\sigma_{v} \mid L/2}: \qquad \pqty*{r,\phi,z}\mapsto  \pqty*{r,-\phi,z+L/2}  
    \label{eqn:glide}
\end{equation}
Repeated application of $Z$ generates the full periodic structure from the monomer. This procedure guarantees exact symmetry consistency by construction.

\subsubsection{Creation of \ce{MoS2}-type nanotubes based on chirality}
For hexagonal two-dimensional lattices, a widely used approach to generate quasi-1D structures is the chiral roll-up construction, which is fully characterized by a pair of integer chiral indices $(n,m)$. For a single-layer hexagonal lattice, several mature software packages support this construction, such as \textsc{ASE} \cite{ase-paper} and \textsc{TubeGen online} \cite{tubegen}. 
However, for more general \ce{MX2}-type, easy-to-use and robust structure-generation tools are still lacking. Here we present a dedicated tool that enables the straightforward generation of single- and multilayer \ce{MoS2}-type nanotubes directly from specified chiral indices.

\begin{figure}[H]
    \centering
    \includegraphics[width=.8\linewidth]{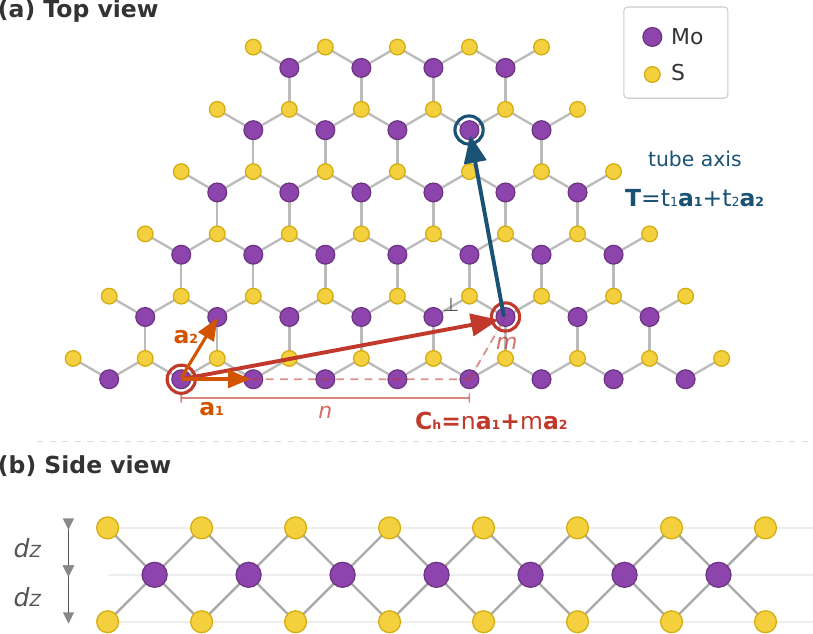}
    \caption{Geometric definition of chiral indices $(n,m)$ for an \ce{MoS2}-type nanotube. (a) Top view of the two-dimensional hexagonal parent lattice, showing the primitive lattice vectors $\bm{a}_{1}$ and $\bm{a}_2$, the chiral vector $\bm{C}_h = n\bm{a}_1 + m\bm{a}_2$, and the translational vector $\bm{T}$ parallel to the nanotube axis. (b) Side view of the \ce{S-Mo-S} sandwich structure, showing the interlayer spacing $dz$ between the \ce{Mo} layer and each \ce{S} layer.}
    \label{fig:hex}
\end{figure}

The roll-up procedure implemented in \textsc{Pulgon-tools} consists of the following steps:

\paragraph{(a) Specification of chiral indices} 
The chiral vector $\bm{C}_{h} = n\bm{a}_{1} + m\bm{a}_{2}$ is constructed from the primitive lattice vectors $\bm{a_{1}}$ and $\bm{a}_{2}$ of the two-dimensional hexagonal lattice (see Fig.~\ref{fig:hex}), where $a=\vqty*{\bm{a}_{1}}=\vqty*{\bm{a}_{2}}$ is the in-plane lattice constant. The chiral vector defines both the circumference of the resulting nanotube and the orientation of the tube axis.

\paragraph{(b) Helical line-group analysis}
The nanotube radius is determined directly by the length of the chiral vector:
\begin{equation}
    R=\frac{\vqty*{\bm{C}_{_{h}}}}{2\pi} = \frac{a\sqrt{n^{2}+nm+m^{2}}}{2\pi}. 
    \label{eqn:R}
\end{equation}
Let $d = \mathrm{gcd}(n,m)$, and define the reduced indices $\tilde{n}= n/d$ and $\tilde{m} = m/d$. Furthermore, define $d_{R} = \mathrm{gcd}(2\tilde{m}+\tilde{n}, 2\tilde{n}+\tilde{m})$. Then the integer coordinates of the primitive translation vector $\bm{T}=t_{1}\bm{a_{1}}+t_{2}\bm{a_{2}}$ are given by:
\begin{equation}
    t_1=-\frac{2\tilde{m}+\tilde{n}}{d_{R}}, \quad t_2=\frac{2\tilde{n}+\tilde{m}}{d_{R}}.
\end{equation}
The vector $\bm{T}$ is orthogonal to the chiral vector $\bm{C}_{h}$, and its magnitude $L=\vqty*{\bm{T}}$ defines the shortest axial period consistent with the translational invariance of the structure.

The total number of helical steps per axial period is:
\begin{equation}
    \tilde{q}=\tilde{n}t_{2}-\tilde{m}t_{1}, \qquad Q=d\tilde{q}, \qquad   f=\frac{L}{\tilde{q}}
\end{equation}

The nanotube additionally possesses a pure $d$-fold rotational symmetry $C_{d}$. The generalized translational generator is therefore the screw operation $\pqty*{C_{Q} \mid f}$, the axial point group contains the pure rotation $C_{d}$, and the nanotube radius is given by Eq.~\eqref{eqn:R}. More detailed analysis can be found in the paper by Domnin et al.\cite{nano13192699}.

\paragraph{(c) Mapping to cylindrical coordinates} 
Atomic positions of the two-dimensional parent structure are transformed from Cartesian coordinates $\bm{\rho}=\pqty*{x,y}$ to cylindrical coordinates $(r,\phi,z)$ according to
\begin{equation}
    r = R,\quad   \phi = \frac{2\pi\pqty*{\bm{\rho}\cdot \bm{C}_{h}}}{\vqty*{\bm{C}_{h}}^2} , \quad z=\frac{\bm{\rho}\cdot \bm{T}}{\vqty*{\bm{T}}} .
\end{equation}

\paragraph{(d) Construction of the symmetry cell (symcell)}
The symcell is the minimal set of atoms that generates the complete nanotube when all symmetry operations of the line group are applied to it. For an \ce{MoS2}-type nanotube, the symcell contains one transition-metal atom and two chalcogenide atoms. Here, the initialization selects three atoms connected together.

\paragraph{(e) Bond-length adjustment via constrained optimization} The geometric roll-up of a planar structure onto a cylinder generally distorts bond lengths. In the 2D \ce{MX2} monolayer, the \ce{M} and \ce{X} atoms are separated by an interlayer distance $dz$ in the out-of-plane direction prior to roll-up. After mapping to the cylinder, the two \ce{X} sublayers are placed at radii $r+dz$ and $r-dz$ respectively, where $r$ is the radius of the $M$ sublayer.

To promote structures closer to structural stability, the following conditions are enforced, requiring the distances between each chalcogenide atom $S_{k}$ and its three nearest-neighbor transition-metal atoms $M_{k1},M_{k2},M_{k3}$ to equal the bond length $d_{0}$ of \ce{M-X} in the 2D monolayer structure:
\begin{equation}
 \left\{\begin{matrix}
 \vqty*{\bm{r}_{S_{k}}-\bm{r}_{M_{k1}}}=\vqty*{\bm{r}_{S_{k}}-\bm{r}_{M_{k2}}} \\
  \vqty*{\bm{r}_{S_{k}}-\bm{r}_{M_{k1}}}=\vqty*{\bm{r}_{S_{k}}-\bm{r}_{M_{k3}}} \\
 \vqty*{\bm{r}_{S_{k}}-\bm{r}_{M_{k1}}}=d_{0}
\end{matrix}\right.
\end{equation}

\noindent This nonlinear system of three equations is solved using a Newton-Krylov iterative method as implemented in \textsc{SciPy} \cite{Virtanen2020}, with a zero initial guess.

\paragraph{(f) Group closure and structure generation}
The complete set of symmetry operations is obtained by group closure of the generators and applied to the symcell to generate the full nanotube.

This approach ensures that the generated nanotubes are both geometrically consistent and fully compatible with the underlying line-group symmetry.

\subsection{Line-group symmetry detection}
\label{sec:sym_detection}
Starting from an atomistic input structure, the symmetry-detection algorithm identifies generalized translational symmetries and axial-point-group operations that are characteristic of quasi-1D systems, and combines them to construct the full line group. A line group can be expressed as the semidirect product of a generalized translational group $Z$ and an axial point group $P$:
\begin{equation}
    L=Z\rtimes P
\end{equation}
which naturally leads to a two-stage symmetry detection strategy. From an implementation perspective, our approach largely builds upon and extends the symmetry-operation framework and structural representations provided by \textsc{pymatgen} \cite{pymatgen}.

\subsubsection{Generalized translational group detection}
The generalized translational group describes symmetry operations that combine translations along the periodic axis with possible screw rotation or glide reflection.

\paragraph{(a) Identification of the periodic direction} 
The input structure is assumed to be periodic along the Cartesian $z$ axis. The algorithm first verifies the existence of translational periodicity along this direction and projects all atomic coordinates onto the axial coordinate for subsequent analysis.

\paragraph{(b) Determination of the central axis} 
The center of mass of the structure is computed and used to locate the position of the central axis in the OXY plane. This central axis is used in the subsequent screw rotation and axial point group rotation detection.

\paragraph{(c) Primitive translation detection}
The primitive translational period $L$ along the $z$ axis is determined by searching for the smallest pure translation that maps the structure onto itself within a prescribed numerical tolerance. This pure translation defines the primitive one-dimensional lattice of the system.

\paragraph{(d) Monomer identification}
Based on the primitive translation $L$, the algorithm extracts a set of candidate monomers, defined as the non-redundant atomic groups that generate the full structure under generalized translational operations. Each monomer is associated with a corresponding translation distance $f\in(0, L]$ and is tested independently in the subsequent step.

\paragraph{(e) Screw-axis detection} 
The program first identifies potential helical rotation angles, defined by a rotational order $Q$, and then, combining them with a translation $f$, detects the helical symmetry operations of the structure. The generalized translation is denoted as Eq.~\ref{eqn:screw}.

\noindent where $C_{Q}$ corresponds to an $Q$-fold rotation about the axis and $f$ is the fractional translation along $z$ mentioned above.

\paragraph{(f) Glide-symmetry detection} 
The presence of mirror symmetry planes containing the $z$ axis is tested. If such a symmetry exists in combination with a translation, that generalized translation is classified as a glide operation described as Eq.~\ref{eqn:glide}.
A structure may simultaneously exhibit screw-axes and glide symmetries, and both are detected independently.

\subsubsection{Axial-point-group detection}
The axial point group is determined by analyzing the primitive cell as a rigid cluster, without imposing periodic boundary conditions. The identification proceeds through the following steps.

\paragraph{(g) Determination of the principal rotational symmetry}
The algorithm searches for the highest-order $l$-fold rotation symmetry $C_l$ about the $z$ axis. Candidate rotation angles are:
\begin{equation}
    \theta_{k} = \frac{2\pi}{k}, \quad k=2,3,...,k_{max}
\end{equation}
where $k_{max}$ is set to the minimum cardinal among all sets of atoms sharing the same $z$ coordinate, since the rotational axis $z$ has already been determined. For each candidate $k$, the algorithm verifies that rotating every atom by $\theta_{k}$ about the $z$ axis yields a position occupied by an atom of the same species within tolerance. The largest valid $k$ defines $l$.

\paragraph{(h) Detection of perpendicular twofold rotations}
If a nontrivial rotational symmetry is present ($l>1$), the algorithm tests for the existence of twofold axes ($U_{d}$ operations) perpendicular to $z$:
\begin{equation}
    U_{d}: \pqty*{\phi, r, z}  \mapsto  \pqty*{\phi_{0}-\phi,r,-z}
\end{equation}
where the azimuthal angle $\phi_{0}$ depends on the twofold axis $d$. The presence of $U_{d}$ symmetry indicates a dihedral-type axial point group.

\paragraph{(i) Classification of dihedral point groups}
When $U_{d}$ symmetry is detected, the axial point group is classified as either $D_{nh}$ or $D_{nd}$, depending on the additional presence of horizontal mirror planes or improper rotation symmetries, respectively.

\paragraph{(j) Classification of cyclic and improper point groups}
If $U_{d}$ symmetry is absent but $l>1$, the point group is classified as $C_{nh}$, $C_{nv}$, or $S_{2n}$ according to whether a horizontal mirror plane, a vertical mirror, or an improper rotation is detected.

\paragraph{(k) Low-symmetry cases}
If no nontrivial rotational symmetry exists ($l=1$), the axial point group reduces to  $C_s$ (mirror plane only), $C_i$ (inversion only), or $C_1$ (no symmetry).

\subsection{Matrix representations and the character table}
\label{subsec:irrep}
Irreducible representations and their associated character tables play a central role in the symmetry analysis of crystalline systems. They provide the theoretical foundation for the classification of electronic band structures and phonon modes \cite{reich2002carbon}, the derivation of symmetry-based selection rules \cite{PhysRevB.65.045418}. For quasi-1D systems, these tasks require the use of line-group representations.

A complete and systematic construction of irreducible representation matrices and character tables for all $13$ families of line groups, see Table~\ref{tab:line_group}, was established by Damnjanovi\'c and Milo\v{s}evi\'c \cite{M_Vujicic_1977,Bozovic_1981,damnjanovic2010line}. In their framework, each irrep is uniquely labeled by a set of physically meaningful quantum numbers: the axial wave vector $k\in (-\pi/L, \pi/L]$, the angular momentum index $m$ and parity labels $\Pi_{V}$, $\Pi_{H}$, $\Pi_{U}$ associated with vertical, horizontal mirror and twofold ($U$-axis) rotational symmetries respectively. 
The allowed values of $m$ and the presence of parity labels depend on the specific line-group family: families containing a vertical mirror symmetry (e.g., families 6-13) admit a $\Pi_{V}$ label; families containing a horizontal mirror symmetry (e.g., families 2-4,9-13) admit a parity label $\Pi_{H}$; and families containing an improper rotational symmetry (e.g., families 5, 9-13) admit a parity label $\Pi_{U}$. Families without any such symmetry (e.g., family 1) are labeled by $(k,m)$ alone.

In \textsc{Pulgon-tools}, the irrep framework of Damnjanovi\'c and Milo\v{s}evi\'c \cite{damnjanovic2010line} is implemented analytically, without reliance on precomputed lookup tables. Starting from the line group identified by the symmetry detection module (Sec.~\ref{sec:sym_detection}), the code determines the family index and constructs the full list of quantum numbers $\left( k,m,\Pi_{V},\Pi_{H},\Pi_{U} \right)$ for the requested wave vector $k$. For each irrep, the full representation matrices $\bm{D}^{(k,m,\Pi)}(g)$ are constructed for every symmetry operation $g$ of the group, and the characters are obtained as their traces:
\begin{equation}
    \chi^{(k,m,\Pi)}(g)=\mathrm{Tr}[\bm{D}^{(k,m,\Pi)}(g)] .
\end{equation}
The choice of quantum number convention $(k,m)$ is discussed in Section~\ref{subsec:irreps}.

\subsection{Force-constant correction}

\textsc{Pulgon-tools} includes a dedicated module for enforcing physical invariance conditions on second-order IFCs. This functionality is designed as a post-processing step for force constants obtained from first-principles calculations or machine-learning interatomic potentials. The relevance of these invariances is even greater in subperiodic (quasi-1D and quasi-2D) systems than in regular 3D crystals.

\subsubsection{Physical constraints on harmonic force constants}
Within the harmonic approximation, the second-order IFCs defining the quadratic potential energy surface must satisfy fundamental physical invariance principles, including conservation of linear and angular momentum. 
These requirements lead to a hierarchy of linear constraints, commonly referred to as translational acoustic sum rules, Born–Huang rotational sum rules, and Huang invariance conditions \cite{10.1119/1.1934059, Lin2022}.

Violations of these constraints, which frequently arise due to errors, numerical noise, approximate boundary conditions, finite supercell sizes, or incomplete symmetry enforcement, can lead to unphysical phonon behavior, such as spurious finite frequencies of acoustic modes at the Brillouin-zone center or incorrect low-frequency dispersions in subperiodic systems \cite{Cen2026, Carrete01102016}.

In the following, $\bm{\Phi}_{ij}^{\alpha \beta}$ denotes the IFC coupling the Cartesian component $\alpha$ and $\beta$ of atoms $i$ and $j$, respectively, and $\bm{r}_{ij}^{\gamma}$ denotes the $\gamma$ Cartesian component of the displacement vector from atom $i$ to atom $j$ under the minimum-image convention.

The translational acoustic sum rules require that
\begin{equation}
    \sum_{i}\bm{\Phi}_{ij}^{\alpha \beta}=0
\end{equation}
while rotational invariance demands that
\begin{equation}
    \sum_{j}\bm{\Phi}_{ij}^{\alpha \beta}\bm{r}_{j}^{\gamma}=\sum_{j}\bm{\Phi}_{ij}^{\alpha \gamma}\bm{r}_{j}^{\beta},
\end{equation}
and the Huang invariance conditions constrain the second moments of the force constants:
\begin{equation}
    \sum_{ij}\bm{\Phi}_{ij}^{\alpha \beta}\bm{r}_{ij}^{\gamma}\bm{r}_{ij}^{\delta}=\sum_{ij}\bm{\Phi}_{ij}^{ \gamma\delta}\bm{r}_{ij}^{\alpha}\bm{r}_{ij}^{\beta}.
\end{equation}

To maintain consistency in systems with periodic boundary conditions, we specifically enforce a transpose symmetry between the adjacent cells (cell $1$ and cell $N-1$) and the first cell (cell $0$): 
\begin{equation}
    \bm{\Phi}_{i(0)j(1)}^{\alpha \beta} = \bm{\Phi}_{j(0)i(N-1)}^{\beta\alpha}
\end{equation}

Finally, to ensure the IFCs remain physically short-ranged and reduce the computational cost, we restrict the interaction of atoms within the adjacent cells: 

\begin{equation}
\bm{\Phi}_{i(N_{1})j(N_{2})}^{\alpha \beta} = 0, \quad \text{when} \; N_{2} \notin \left\lbrace N_{1}-1, N_{1}, N_{1}+1\right\rbrace
\end{equation}
This constraint prevents the optimization from redistributing residuals from spurious violations of the symmetries among unphysical long-range interactions.

\subsubsection{Constraint construction and optimization strategy}
All invariance conditions are formulated as a single sparse linear constraint system 

\begin{equation}
    \bm{C} \cdot \bm{H} = 0,
\end{equation}
where $\bm{H}$ is the harmonic IFC tensor \enquote{flattened} into a single vector and $\bm{C}$ is a sparse matrix containing the coefficients of all imposed constraints.

To enforce the full set of constraints while keeping the perturbation to the original force constants small, the correction is formulated as a linearly constrained quadratic optimization problem,

\begin{equation}
\bm{H}_{\mathrm{corrected}}
=
\argmin_{\bm{H}}
\left\|
\bm{H} - \bm{H}_{\mathrm{raw}}
\right\|^{2}
\end{equation}
where $\bm{H}_{\mathrm{raw}}$ and $\bm{H}_{\mathrm{corrected}}$ denote the uncorrected and corrected IFCs tensor, respectively.

This formulation ensures that the corrected force constants satisfy all physical invariance conditions while deviating as little as possible from the original numerical data. The linear constraint system is typically highly overdetermined and sparse, reflecting the large number of invariance relations compared to the number of independent IFC elements.

The constrained optimization problem is solved using the convex optimization framework \textsc{CVXPY} \cite{cvxpy}, with the OSQP solver \cite{osqp} employed for efficient handling of large sparse systems. As an alternative option, \textsc{Pulgon-tools} also has a backend based on \textsc{sklearn}'s implementation of ridge regularization \cite{pedregosa2011scikit} for enforcing the constraints, which may be useful in memory-limited scenarios.

\section{Examples of application}
\label{sec:examp}
To demonstrate the practical usefulness of \textsc{Pulgon-tools}, we present a series of illustrative applications covering all modules. The examples selected illustrate different aspects of the code and validate its performance for realistic quasi-1D systems.

\subsection{Generation of quasi-1D structure}
\subsubsection{The generation process of the General Symmetry-based approach}

\begin{figure}[H]
    \centering
    \includegraphics[width=1\linewidth]{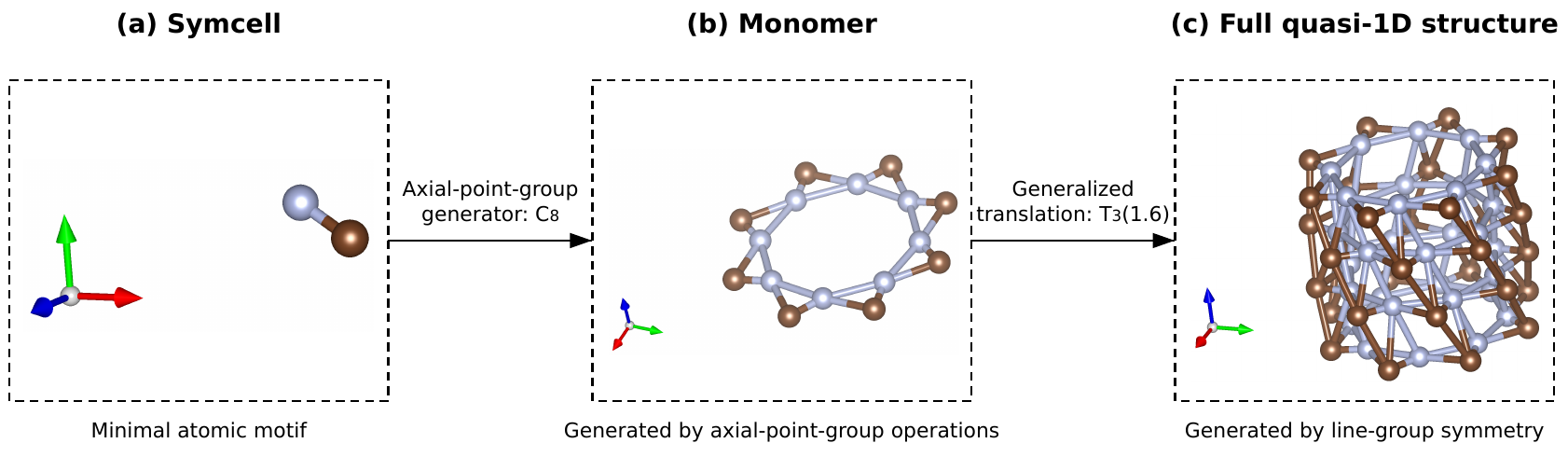}
    \caption{Illustration of the general symmetry-based construction of a quasi-1D structure. (a) A minimal set of atomic positions is specified in cylindrical coordinates. (b) The monomer is generated by applying the axial-point-group operations defined by the line group. (c) The full quasi-1D structure is obtained by applying the generalized translational symmetry, combining rotation and fractional translation along the symmetry axis.}
    \label{fig:sym-based}
\end{figure}

Quasi-1D structures with arbitrary line-group symmetry can be generated directly from motif and symmetry generators using:
\begin{pycode}
pulgon-generate-structures-sym_based \
-m R1 PHI1 Z1 \
   R2 PHI2 Z2 \
-b SYMBOL \
-g APG_GENERATOR \ 
-c TG_GENERATOR \
-s OUTPUT_FILENAME
\end{pycode}

For the representative example shown in Fig.~\ref{fig:sym-based}, the input parameters are:
\begin{pycode}
R1 PHI1 Z1 R2 PHI2 Z2 = 3 0 0 2.2 0.2618 0
SYMBOL = C N
APG_GENERATOR = "Cn(8)"
TG_GENERATOR = T_Q 3 1.6
OUTPUT_FILENAME = structure.vasp
\end{pycode}
This corresponds to a system with eightfold rotational symmetry $C_{8}$ and a screw-type generalized translation with rotational number $Q=3$ and translation $f=1.6 $ \AA. 

\subsubsection{\ce{MoS2}-type nanotubes from chiral indices}
\begin{figure}[H]
    \centering
    \includegraphics[width=1\linewidth]{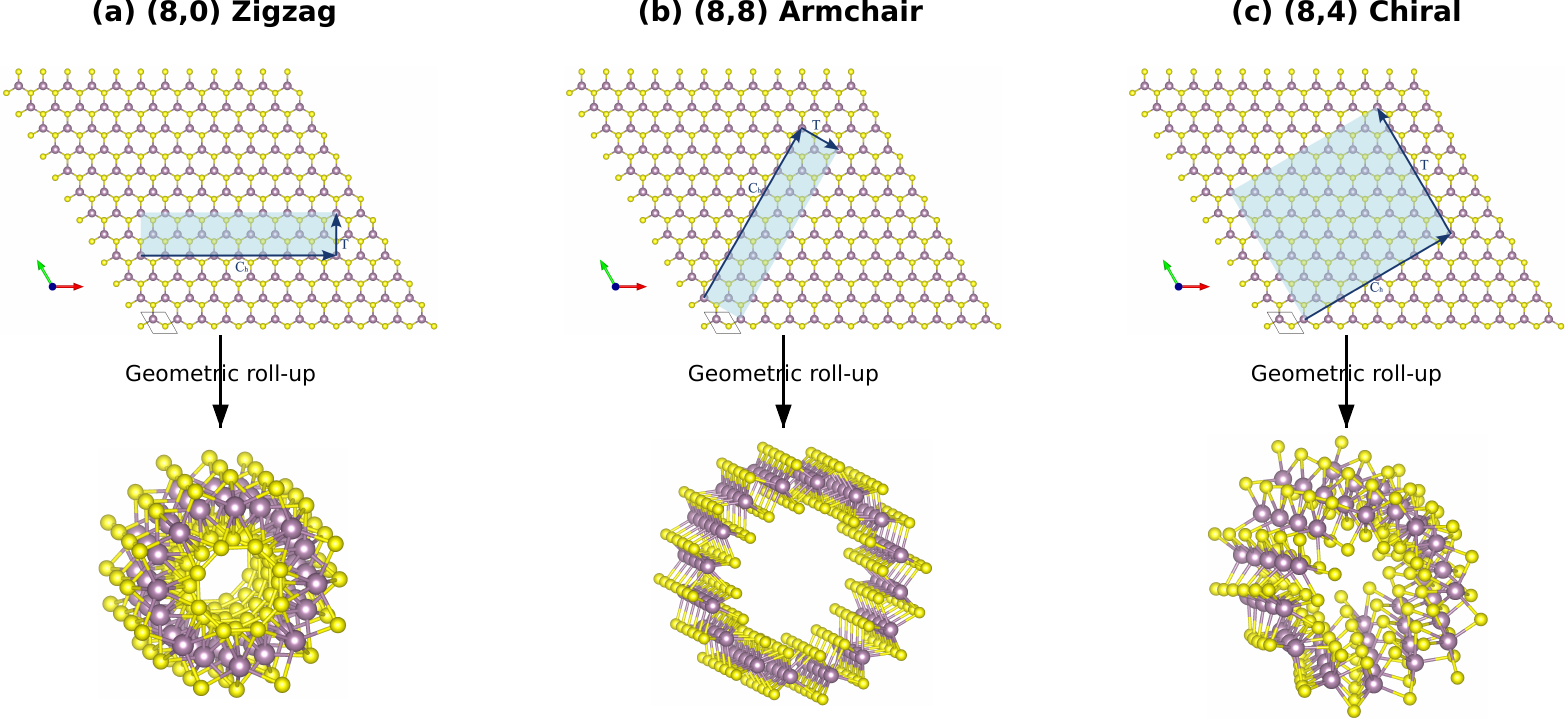}
    \caption{Illustration of the chiral roll-up construction of quasi-1D nanotubes from a two-dimensional hexagonal lattice.
For each set of chiral indices $(n,m)$, the nanotube is generated by geometrically rolling the parent lattice along the chiral vector $    C_{h} = n\bm{a}_{1} + m\bm{a}_{2} $.
Representative examples of armchair $(8,8)$, zigzag $(8,0)$, and chiral $(8,4)$ nanotubes are shown.}
    \label{fig:mos2}
\end{figure}

For \ce{MX2}-type hexagonal parent lattices, nanotubes can be generated via the chiral roll-up approach:
\begin{pycode}
pulgon-generate-structures-chirality \ 
-c CHIRALITY \ 
-b METAL CHALCOGEN \ 
-l BOND_LENGTH \ 
-d DZ \
-s OUTPUT_FILENAME
\end{pycode}

Representative examples are shown in Fig.~\ref{fig:mos2}; for these examples the input parameters are:
\begin{pycode}
CHIRALITY = 8 0 / 8 8 / 8 4
METAL CHALCOGEN = Mo S
BOND_LENGTH = 2.43
DZ = 1.57
OUTPUT_FILENAME = POSCAR
\end{pycode}
These examples demonstrate how structures with different chiralities can be generated easily and stably. Note that the structures cannot be generated for unreasonable chiral indices which would lead to tiny diameters.

\subsection{Symmetry detection of single-wall carbon nanotube}
\begin{figure}[H]
    \centering
    \includegraphics[width=1\linewidth]{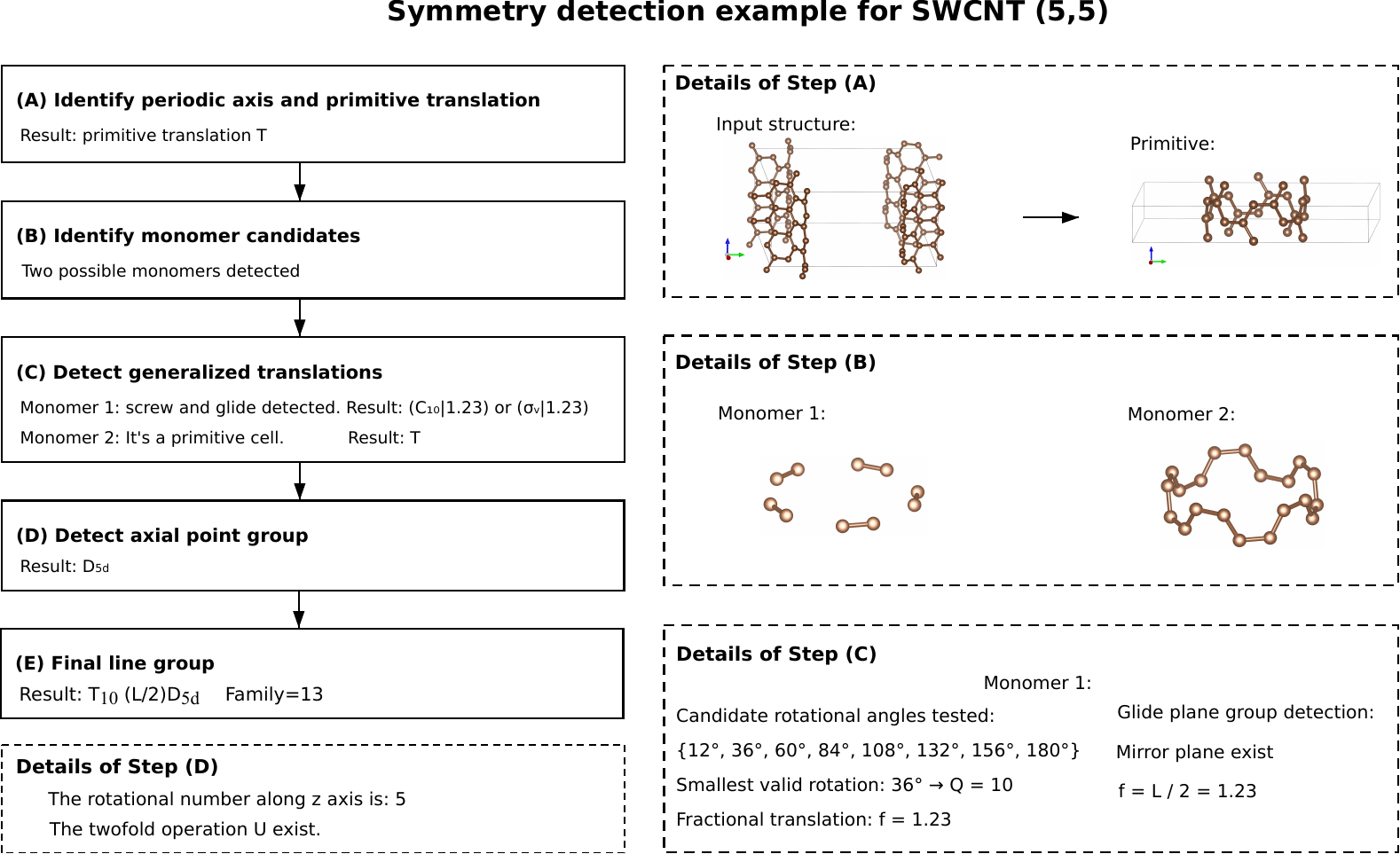}
\caption{Symmetry detection workflow for a $(5,5)$ single-wall carbon nanotube (SWCNT).}
    \label{fig:sym_detection}
\end{figure}

The symmetry detection procedure is executed via:
\begin{pycode}
pulgon-detect-AxialPointGroup -p POSCAR -t 0.01
pulgon-detect-CyclicGroup -p POSCAR -t 0.01 --layer-tolerance 0.05
pulgon-detect-linegroup -p POSCAR -t 0.01 --layer-tolerance 0.05
\end{pycode}

\noindent Here, the tolerance parameter controls the Cartesian coordinate matching threshold used to identify symmetry-equivalent atoms in both the generalized translational group and axial point group detection. The layer-tolerance parameter is used only in the generalized-translational-group detection stage. It defines a fractional tolerance along the periodic $z$ direction for grouping atoms into layers and generating monomer translation candidates. In the examples shown here, the default value $0.05$ is used. An example for detecting the symmetry of single-wall carbon nanotube (SWCNT) is shown in Fig.~\ref{fig:sym_detection}. The workflow highlights the branching logic associated with multiple monomer candidates and the identification of the generalized translational and axial point group symmetries. For this $(5,5)$ SWCNT, two possible monomer candidates are identified (Fig.~\ref{fig:sym_detection}, step B). The algorithm branches and evaluates each candidate independently. For one candidate, a screw operation $\left( C_{10} \mid L/2 \right)$ and a glide operation $\left( \sigma_{v} \mid L/2 \right)$ are detected. The rotational order detected around the periodic axis is five. The final line group is determined as $T_{10}(L/2)D_{5d}$ and belongs to line-group family 13.

\subsection{Irreps and character table}
\begin{figure}[H]
    \centering
    \includegraphics[width=1\linewidth]{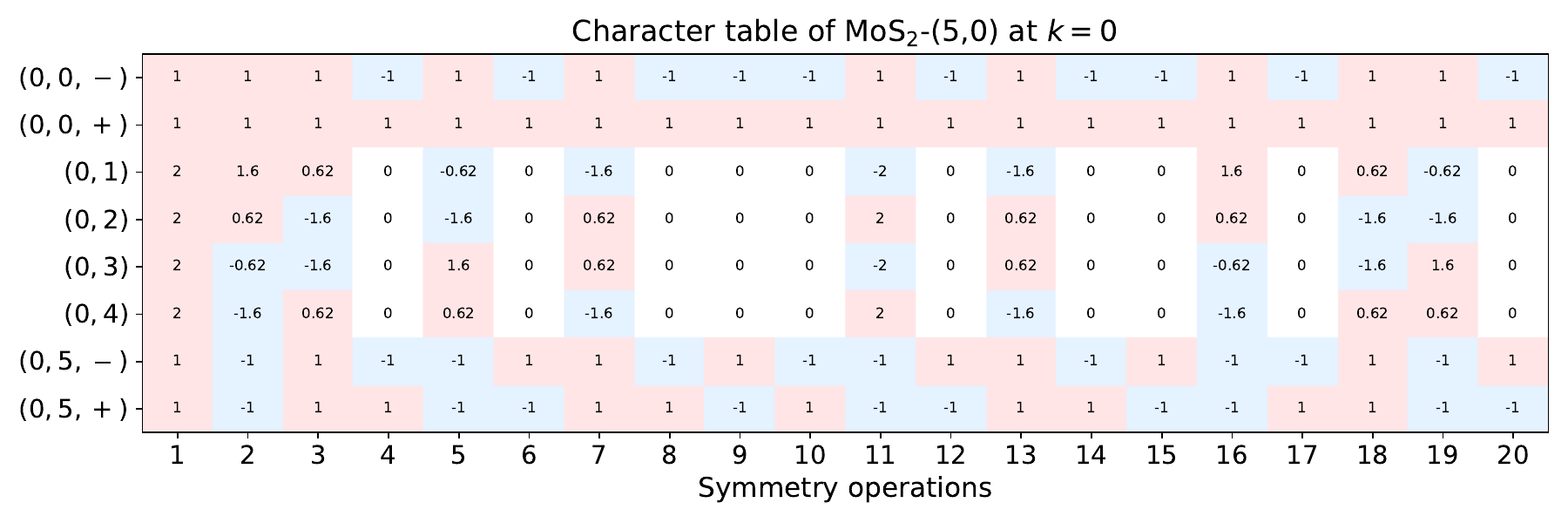}
\caption{Character table of the eight irreducible representations of the \ce{MoS2}-(5,0) nanotube at $k=0$. Rows correspond to irreps labeled by $\left( k,m,\Pi_{V} \right)$, while columns correspond to the 20 symmetry operations of the line group (indexed as operations $1$–$20$). Positive and negative characters are indicated by light red and light blue shading, respectively. The periodic dependence on the angular momentum index $m$ and the mirror-parity splitting for $\Pi_{V}=\pm 1$ are clearly visible.}
    \label{fig:chara}
\end{figure}

The irreps and character table can be generated using:
\begin{pycode}
pulgon-irreps-tables -p POSCAR -q 0.0 -t 0.01 -u 1e-6 -s characters
\end{pycode}
\noindent Here `-t' (`--tolerance') is the Cartesian atomic-position tolerance with units \AA, used in the preliminary line-group symmetry detection. It controls whether symmetry-transformed atoms are matched to atoms in the input structure. The `-u` (`--qpoint-tolerance') option is a dimensionless numerical tolerance used only in the irrep-table construction, for example whether the reduced axial wave vector is equivalent to special values such as $\Gamma$ or the Brillouin-zone boundary. Starting from the detected line group, the code constructs irreps parameterized by axial wave vector $k$, angular momentum index and parity labels. As shown in Fig.~\ref{fig:chara}, for the \ce{MoS2} example, the irreps generated at $k=0$, yielding $8$ irreps and $20$ symmetry operations.

\subsection{IFC correction}

\begin{figure}[H]
    \centering
    \includegraphics[width=.8\linewidth]{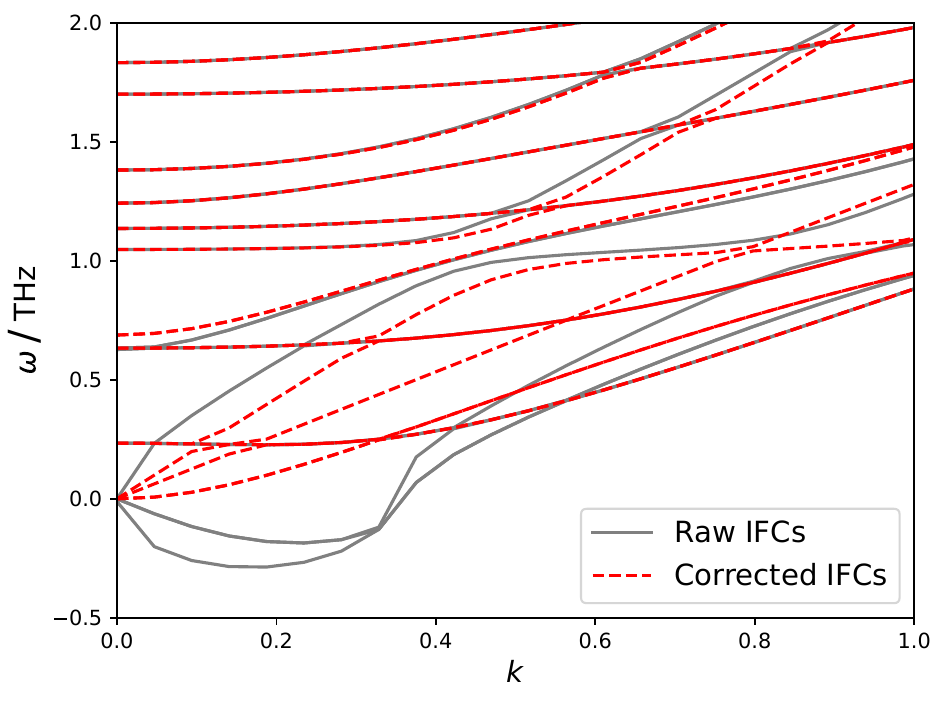}
    \caption{Effect of force constant correction on the phonon dispersion of a (12,12) \ce{MoS2} nanotube. The solid gray lines correspond to the phonon modes near $\Gamma$ that use original IFCs. The red dashed lines correspond to the phonon modes computed from the IFCs correction implemented in \textsc{Pulgon-tools}.}
    \label{fig:fcs_correction}
\end{figure}

The IFC correction is performed via:
\begin{pycode}
pulgon-fcs-correction \ 
-p POSCAR \ 
-x "[1,1,3]" \ 
-f FORCE_CONSTANTS \ 
-m convex_opt \
-n
\end{pycode}

As shown in Fig.~\ref{fig:fcs_correction}, the IFCs correction module is tested on a $(12,12)$ \ce{MoS2} nanotube. The raw second-order IFCs are generated by \textsc{Phonopy}~\cite{phonopy-phono3py-JPCM, phonopy-phono3py-JPSJ} from machine-learning interatomic potentials developed for Ref.~\citenum{Cen2026}. The figure compares the phonon spectra computed from original IFCs and symmetry-corrected IFCs. Before the correction, spurious imaginary frequencies appear near $\Gamma$ due to the violations of translational and rotational invariance. After applying the invariance constraints, all imaginary frequencies are eliminated, restoring the linear and quadratic acoustic branches. 

\subsection{Python API example: IFC correction in combination with Phonopy}
Beyond the command-line interface, \textsc{Pulgon-tools} can also be used directly as a Python library, enabling integration into custom workflows. 
We present a minimal working example that combines \textsc{Phonopy} \cite{phonopy-phono3py-JPSJ, phonopy-phono3py-JPCM} with \textsc{Pulgon-tools} to perform IFC correction.

\begin{pycode}
import numpy as np
import phonopy
from pulgon_tools.force_constant_correction import (
    build_constraint_matrix, solve_fcs
)

# Step 1: Load structure with Phonopy
unitcell = phonopy.interface.vasp.read_vasp("POSCAR")
phonon = phonopy.Phonopy(unitcell, supercell_matrix=np.diag([1, 1, 3]))
phonon.force_constants = phonopy.file_IO.parse_FORCE_CONSTANTS(
    "FORCE_CONSTANTS"
)

# Step 2: Build the sparse constraint matrix that enforces all constraints
M, IFC = build_constraint_matrix(phonon, cut_off=15.0, pbc=[False, False, True])

# Step 3: Solve the constrained quadratic optimization problem
IFC_corrected = solve_fcs(IFC, M, methods="convex_opt")

# Step 4: Save the corrected IFCs
phonopy.file_IO.write_force_constants_to_hdf5(
    IFC_corrected, filename="FORCE_CONSTANTS_correction.hdf5"
)

\end{pycode}

More generally, access to the Python API allows users to build custom analysis pipelines by combining \textsc{Pulgon-tools} modules programmatically. All four modules expose a Python API that can be imported and called directly, enabling fine-grained control over intermediate results and integration with other libraries. Additional usage examples covering all modules are provided in the \texttt{examples} directory of the project repository.

\section{Summary and Conclusions}
\label{sec:sum}
In this work we have presented \textsc{Pulgon-tools}, an open-source software package for the automatic symmetry analysis of quasi-1D periodic systems based on line-group theory. The code provides a robust and systematic framework for identifying the full line group symmetry directly from atomistic structures, even in the presence of numerical distortions.

We described in detail the symmetry detection algorithm implemented in \textsc{Pulgon-tools}, which decomposes the line group identification problem into generalized translational group detection and axial point group analysis. Representative examples, including single-wall carbon nanotubes, demonstrate the ability of the algorithm to resolve multiple monomer candidates and correctly identify the full symmetry group.

Beyond symmetry detection, \textsc{Pulgon-tools} provides a set of auxiliary functionalities tailored to symmetry-aware modeling of quasi-1D materials. These include two complementary approaches for structure generation: a general symmetry-based construction applicable to arbitrary line groups and a dedicated chiral roll-up method for \ce{MoS2}-type nanotubes. In addition, the force constant correction module enforces translational and rotational invariance conditions on harmonic interatomic force constants. The integration of irreducible representations and character tables further facilitates symmetry-resolved analyses of electronic and vibrational properties.

Taken together, \textsc{Pulgon-tools} fills the gap in the detection and analysis of the line group symmetry for quasi-1D materials. By combining automated symmetry detection with structure generation, IFCs post-processing, and representation theory utilities, the package provides a versatile and practical platform for modeling and analysis of quasi-1D systems.

\section*{Data availability}
The latest version of \textsc{Pulgon-tools} is available on \href{https://github.com/pulgon-project/pulgon_tools/releases}{\url{https://github.com/pulgon-project/pulgon_tools/releases}}. The version used for this manuscript is registered on Zenodo with \href{https://zenodo.org/records/21097375}{\url{https://zenodo.org/records/21097375}}.

\section*{Acknowledgements}
 This research was funded in whole or in part by the Austrian Science Fund (FWF) [10.55776/P36129]. For open access purposes, the author has applied a CC BY public copyright license to any author-accepted manuscript version arising from this submission. It was also supported by MCIN with funding from the European Union NextGenerationEU (PRTR-C17.I1) promoted by the Government of Aragon. J.C. acknowledges Grant CEX2023-001286-S funded by MICIU/AEI /10.13039/501100011033.






\bibliographystyle{elsarticle-num}
\bibliography{cpc-cpip}







\end{document}